# Investigation of Crystal Structure of SrLa(FeTi)O$_6$ and BaLa(FeTi)O$_6$ Perovskites by Rietveld Refinement

Uma Shankar, Puneet Kumar Agarwal, Rishikesh Pandey and Akhilesh Kumar Singh*

*School of Materials Science and Technology, Indian Institute of Technology, Banaras Hindu University, Varanasi- 221005*

***Email:** akhilesh_bhu@yahoo.com

The structures of the recently reported ordered perovskites SrLa(FeTi)O$_6$ and BaLa(FeTi)O$_6$ have been revisited using Rietveld analysis of the powder x-ray diffraction data. Our studies clearly show that earlier authors have incorrectly reported the tetragonal structure in the *I4/m* space group for SrLa(FeTi)O$_6$ and the cubic structure in the *Fm$\bar{3}$m* space group for BaLa(FeTi)O$_6$. The correct structure is found to be orthorhombic in the *Pnma* space group for SrLa(FeTi)O$_6$ while cubic in the *Pm$\bar{3}$m* space group for BaLa(FeTi)O$_6$. Thus the BaLa(FeTi)O$_6$ is not an ordered double perovskite as the occupancies of Ba/La ions at A-site and Fe/Ti ions at B-site are disordered resulting in the primitive cubic unit cell in the *Pm$\bar{3}$m* space group. Similarly, the occupancies of Sr/La ions at A-site and Fe/Ti ions at B-site are disordered in SrLa(FeTi)O$_6$ also, ruling out the ordered perovskite structure.

# 1. Introduction

The ideal perovskite structure with stoichiometry $ABO_3$, where A is typically a large, low oxidation state cation, B is a smaller transition metal or lanthanide cation, has cubic symmetry with space group $Pm\bar{3}m$ [Mitchell, 2002]. The beauty of the perovskite structure is that it can accommodate variety of ions at A and B-sites with different valance and ionic sizes leading to widely different physical properties. The Complex perovskite oxide materials are special in their ability to possess variety of widely different physical properties such as ferromagnetic, ferroelectric, piezoelectric, electro-optic, electrocatalytic, insulating, semiconducting, metallic etc. They also exhibit interesting phase transitions involving very rich physics and chemistry. Because of the wide range of physical properties, the multifunctional perovskite oxide solid solutions offer countless applications such as piezoelectric transducers and actuators, magnetic and ferroelectric memories, materials for electro-optic devices and optical fibers, Positive temperature coefficient of resistance materials, [Uchino, 2000, Eerenstein et al., 2006], Colossal magneto resistance materials [Coey et al., 1999], hybrid solar and fuel cell materials [Minh, 1993], phosphors and display devices, sensors etc. The physical properties of these materials are strongly correlated with the crystal structures as well as crystallographic phase transitions. The coexistence of two or more crystallographic phases is observed in these materials around phase transition and in some cases plays very vital role in deciding the physical properties. The perovskite oxides are therefore being explored extensively for structure property correlations.

In complex perovskites two (or more) cations of different valence may accommodate the 'A' and 'B' sites which can significantly affect the physical behaviour and crystal structure. For such systems it is known that when the charge and/or size difference is large for the A or B-site cations, ordering of ions can occur at the A and/or B-sites [Knapp & Woodward et al., 2006].

The ordering often results in the double perovskite structure with the doubling of the primitive perovskite unit cell. To mention, examples of such double perovskite materials are NaLaMgWO$_6$ [King et al., 2009], SrLaCoRuO$_6$ [Kim et al., 1995 & Bos et al., 2005], SrLaZnRuO$_6$, SrLaMgRuO$_6$, SrLaNdRuO$_6$ [Iturbe-Zabalo, Faik et al., 2013], SrPrMgRuO$_6$ [Iturbe-Zabalo, Igartua et al., 2013], SrLaZnRuO6, SrLaMgRuO6 [ Iturbe-Zabalo, Gateshki et al., 2013], SrLaCuNbO$_6$, SrLaCuTaO$_6$ [West & Davies 2012], family of SrLnFeRuO$_6$, where Ln are rare earth ions [ Iturbe-Zabalo, et al., 2012], Sr$_2$FeReO$_6$ [Kobayashi et al., 1999], Sr$_2$FeMoO$_6$ [Kobayashi et al., 2006], Sr$_2$YTaO$_6$ and Sr$_2$YNbO$_6$ [Howard et al., 2005], Sm$_2$NaIrO$_6$ [Samuel et al., 2005], Sr$_2$CaWO$_6$ [Madariaga et al., 2010]. These materials have been extensively studied due to their interesting structural, magnetic, electrical, electronic properties and applications.

The perovskites with magnetic ions at B-sites have drawn considerable attention due to their interesting magnetoresistance behavior and potential for magnetoresistive devices [Kobayashi et al., 1999, Kobayashi et al., 2006]. In the present work we have revisited the structure of recently investigated perovskites SrLa(FeTi)O$_6$ and BaLa(FeTi)O$_6$, the structures of which were incorrectly reported as tetragonal (space group *I*4/*m*) and cubic (space group *Fm$\bar{3}$m*) respectively [Elbadawi et al, 2013]. We show that the correct structure of SrLa(FeTi)O$_6$ is orthorhombic in the *Pnma* space group and that of BaLa(FeTi)O$_6$ is cubic in the *Pm$\bar{3}$m* space group. The results of bond lengths, bond angles and bond valance sum calculations are also reported.

## 2. Experimental

Samples were prepared by conventional solid state ceramic method. Stoichiometric amounts of analytical reagent (AR) grade BaCO$_3$ (HIMEDIA, 99%), Fe$_2$O$_3$ (HIMEDIA, 99%), TiO$_2$ (HIMEDIA, 99%), SrCO$_3$ (HIMEDIA, 99%), La$_2$O$_3$.H$_2$O (HIMEDIA, 99.9%) were used as raw materials. The powders were mixed by ball milling in acetone for 6 h. The mixed powders were dried and then calcined in a muffle furnace for 6 h at 1000 $^0$C, 1200 $^0$C and 1300 $^0$C to optimize the calcination temperature. The calcined powders were checked for phase purity by using 18 kW rotating Cu anode Rigaku (Japan) x-ray diffractometer operating in the Bragg Brentano geometry with the curved crystal graphite monochromator fitted in the diffracted beam. The instrumental resolution is 0.10$^o$. The x-ray diffraction (XRD) data were collected at a scan rate of 2$^o$/min in the 2θ range 20$^0$ to 100$^0$ at the scan step of 0.02$^o$. Rietveld refinements were carried out using the FULLPROF Suite [Carvajal]. The Wyckoff positions and the asymmetric unit for the various space groups used during the Rietveld structure refinement of SrLa(FeTi)O$_6$ are following. For the orthorhombic phase with Pnma space group, we used the lattice parameter a≈a$_p$√2, b≈2a$_p$ and c≈ a$_p$√2 , where a$_p$ corresponds to primitive perovskite unit cell parameter. There are four sites in the asymmetric unit of SrLa(FeTi)O$_6$ with La$^{3+}$/Sr$^{2+}$ ions occupying 4(c) sites at (x, ¼, z), Fe$^{3+}$/Ti$^{4+}$ the 4(a) sites at (0, 0, 0), $O_I^{2-}$ the 4(c) sites at (x, ¼, z), and $O_{II}^{2-}$ the 8(d) sites at (x, y, z). For the cubic phase in the $Pm\bar{3}m$ space group, there are three sites in the asymmetric unit with La$^{3+}$/Ba$^{2+}$ ions occupying the 1(a) sites at (0, 0, 0), Ti$^{4+}$/Fe$^{3+}$ the 1(b) sites at (1/2, 1/2, 1/2), and O$^{2-}$ at the 3(c) sites at (1/2, 1/2, 0). For the $Fm\bar{3}m$ and $I4/m$ space groups the coordinates and lattice parameters were taken from that reported by Elbadawi et al [2013].

## 3. Results and Discussion:

The powder XRD patterns of the SrLa(FeTi)O$_6$ and BaLa(FeTi)O$_6$ ceramics calcined at temperatures 1000 $^0$C, 1200 $^0$C and 1300 $^0$C are shown in Fig.1 and Fig. 2 respectively. We have marked the reflections corresponding to primitive perovskite structure by letter P in the XRD patterns shown in Fig.1 and Fig. 2. It is evident from these figures that solid state reaction is not complete at the calcination temperatures 1000 $^0$C and 1200 $^0$C as there are several peaks marked with asterisks that could not be indexed by considering the perovskite structure. The impurity peaks observed for calcination temperature 1000 $^0$C and 1200 $^0$C are eliminated at the calcination temperature of 1300 $^0$C for SrLa(FeTi)O$_6$. There is still a barely visible reflection around $2\theta \approx 28^0$ in the XRD pattern of BaLa(FeTi)O$_6$ calcined at 1300 $^0$C, which does not correspond to the perovskites phase. We analyzed the XRD patterns of the two samples calcined at 1300 $^0$C using Rietveld structure refinement to resolve the correct structure of these compounds and the results are discussed below.

### 3.1 Structure of SrLa(FeTi)O$_6$:

Elbadawi et al [2013] have reported that the structure of SrLa(FeTi)O$_6$ [SLFT] is tetragonal with space group *I*4*/m*. Rietveld analysis of the XRD data by us reveals that there are several additional super lattice reflections which are left un-indexed by *I*4*/m* space group. Some of such super lattice reflections are shown in Fig. 3 and are marked by arrows. These additional super lattice reflections are present in the XRD data of SLFT reported by Elbadawi et al [2013] also [see Fig.1 of Elbadawi et al., 2013]. This suggests that the consideration of lower symmetry phase is inevitable to account for the structure of SLFT. In view of this, we considered orthorhombic *Pnma* space group corresponding to the tilt system a$^+$b$^-$b$^-$ [Glazer, 1972] to refine the structure of SLFT. The orthorhombic *Pnma* space group is observed in several double perovskite compounds [Knapp & Woodward et al., 2006]. A comparison of the Rietveld fits of the powder XRD data of SLFT considering *I*4*/m* and *Pnma* space groups is shown in Fig. 3. As can be seen from Fig. 3, the super lattice reflections which were not fitted by the *I*4*/m* space group are fitted well by *Pnma* space group. This unambiguously confirms that the structure of SLFT is orthorhombic in the *Pnma* space group. The Rietveld fit of the XRD data of SLFT in the 2θ range 20-100 degrees is shown in Fig. 4. The overall fit is also very good. The refined structural parameters of SLFT for Pnma space group are listed in Table I. Elbadawi et al [2013] have considered ordering of Fe/Ti ions at B-sites while refining the structure of SLFT using $I4/m$ space group. We find that cations at both the A and B-sites are disordered in the structure and account very well the observed XRD pattern of SLFT when *Pnma* space group is used. The unit cell and occupancy of the ions for SLFT is shown in Fig. 5.

### 3.2 Structure of BaLa(FeTi)O$_6$:

Elbadawi et al [2013] have reported that the structure of BaLa(FeTi)O$_6$ [BLFT] is cubic in the $Fm\bar{3}m$ space group. However, examination of the experimental XRD patterns reported by Elbadawi et al and in the present work [see Fig. 2] suggests that there are no super lattice reflections corresponding to the $Fm\bar{3}m$ space group. In Fig. 6, we have shown the Rietveld fits of the selected XRD profiles of BLFT considering the $Fm\bar{3}m$ and $Pm\bar{3}m$ space groups. As can be seen from Fig. 6, there are many super lattice reflections in the calculated pattern for $Fm\bar{3}m$ space group (the 2θ positions are marked by arrows) which are absent in the experimental data. Thus, there is no need to use $Fm\bar{3}m$ space group as the consideration of primitive perovskite unit cell in the $Pm\bar{3}m$ space group accounts very well the experimental data. The structure of BLFT is therefore cubic in $Pm\bar{3}m$ space group and not $Fm\bar{3}m$ reported earlier. The Rietveld fit of the XRD pattern for BLFT in the 2θ range 20 to $100^0$ is shown in Fig. 7. The overall fit is very good confirming the $Pm\bar{3}m$ space group for BLFT. In case of BLFT also, Elbadawi et al [2013] have reported that Fe/Ti ions are ordered at the B-sites. The present work clearly shows that both the A and B-sites cations are disordered in the primitive perovskite structure with $Pm\bar{3}m$ space group. The unit cell and the occupancy of various ions for BLFT is shown in Fig. 8. The refined structural parameters for BLFT are listed in Table II.

## 3.3 Bond Lengths and Bond Angles for SrLa(FeTi)O$_6$ and BaLa(FeTi)O$_6$

Selected bond angles and bond lengths for the SLFT and BLFT are given in Table III. Results of the bond valance sum (BVS) calculations are also listed in this table. The A-O and B-O bond lengths calculated from Shannon ionic radii are 2.80 Å and 1.9775 Å respectively for SLFT. The B-O$_I$ bond lengths obtained from Rietveld structure refinement of SLFT is comparable to that obtained from Shannon ionic radii while B-O$_{II}$ bond lengths are slightly different. In the distorted *Pnma* structure of SLFT, the longest and smallest A-O bond lengths obtained from Rietveld structure refinement are 3.1310(2)Å and 2.3500(4)Å. The bond angles are significantly deviated from 90 and 180$^0$ suggesting that significant orthorhombic distortion is there. The tolerance factor of SLFT calculated from the bond lengths obtained by Rietveld structure refinement is ~0.87. The low value of tolerance factor ~0.87 indicates that the SLFT should crystallize in lower symmetry phase like orthorhombic in *Pnma* space group. The bond valance sum calculations (BVS) for SLFT results in reasonable charges on various ions.

For BLFT, the A-O and B-O bond lengths calculated from Shannon ionic radii are 2.885 Å and 1.9775 Å respectively. The values of the A-O and B-O bond lengths for BLFT calculated from Rietveld structure refinement are comparable from those obtained from Shannon [Shannon, 1976] ionic radii suggesting that the bonding is predominantly ionic in nature. The tolerance factor of BaLaFeTiO$_6$ calculated from the bond lengths obtained from Rietveld structure refinement is ~1.0004 and that from Shannon ionic radii is ~1.0074 [Shannon, 1976]. This further confirms that cubic structure in the *Pm$\bar{3}$m* space group is expected for BLFT. The BVS calculations on BLFT resulted unusually high charge for Ba-ion while significantly low charge on La-ion. This may be due to presence of large local disorder at A-site. The charges on Fe, Ti and O-ions are in the reasonable range.

## 4. Conclusions

Newly reported perovskites SrLa(FeTi)$O_6$ and BaLa(FeTi)$O_6$ have been synthesized by solid-state reaction method. The Rietveld refinement of the structure using x-ray diffraction data reveals that the crystal structure of SrLa(FeTi)$O_6$ and BaLa(FeTi)$O_6$ are orthorhombic (*Pnma* space group) and cubic (*Pm$\bar{3}$m* space group) respectively. We show that the tetragonal (*I4/m*) structure for SrLa(FeTi)$O_6$ and cubic (*Fm$\bar{3}$m*) structure for BaLa(FeTi)$O_6$ reported by Elbadawi et al. [ 2013] is incorrect. Both the perovskites crystallize in disordered structures.

# Figure Captions

Figure 1

Powder XRD patterns of SrLa(FeTi)$O_6$ samples calcined at temperatures 1000°C, 1200°C and 1300°C for 6 hours. The reflections corresponding to impurity phases are marked by asterisks.

Figure 2

Powder XRD patterns of BaLa(FeTi)$O_6$ samples calcined at temperatures 1000°C, 1200°C and 1300°C for 6 hours. The reflections corresponding to impurity phases are marked by asterisks.

Figure 3

Rietveld fit for the selected XRD profiles of SrLa(FeTi)$O_6$ using orthorhombic (*Pnma*) space group (upper panel) and tetragonal (*I4/m*) space group (lower panel). The super lattice reflections are marked by arrows.

Figure 4

Observed (dots), calculated (continuous line) and difference (continuous bottom line) profiles for SrLa(FeTi)$O_6$ obtained after Rietveld analysis of the XRD data using orthorhombic *Pnma* space group. The vertical tick marks above the difference plot show the Bragg peak positions.

Figure 5

The unit cell for SrLa(FeTi)$O_6$ corresponding to the orthorhombic *Pnma* space group.

Figure 6

Rietveld fit for the selected XRD profiles of BaLa(FeTi)$O_6$ using Cubic (*Pm$\bar{3}$m* space group, upper panel) and Cubic (*Fm$\bar{3}$m* space group, lower panel). The arrows mark the positions where super lattice reflections corresponding to Fm$\bar{3}$m space group are expected. The $Cu_{K\alpha 1}$ and $Cu_{K\alpha 2}$ doublets are seen to be clearly resolved.

Figure 7

Observed (dots), calculated (continuous line) and difference (continuous bottom line) profiles for BaLa(FeTi)$O_6$ obtained after Rietveld analysis of the XRD data using cubic *Pm$\bar{3}$m* space group. The vertical tick marks above the difference plot show the Bragg peak positions.

Figure 8

The unit cell for BaLa(FeTi)$O_6$ corresponding to the cubic *Pm$\bar{3}$m* space group.

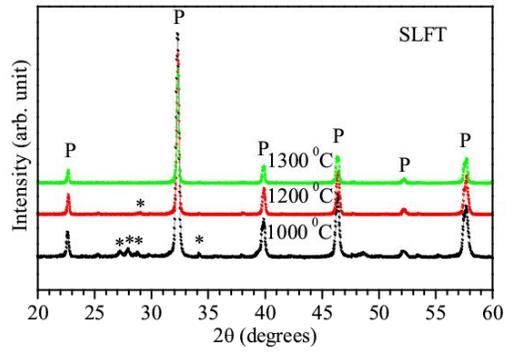

Figure 1

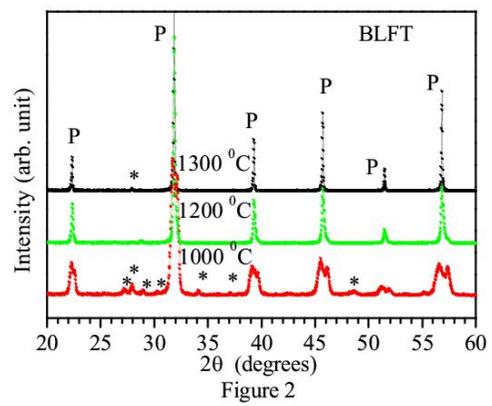

Figure 2

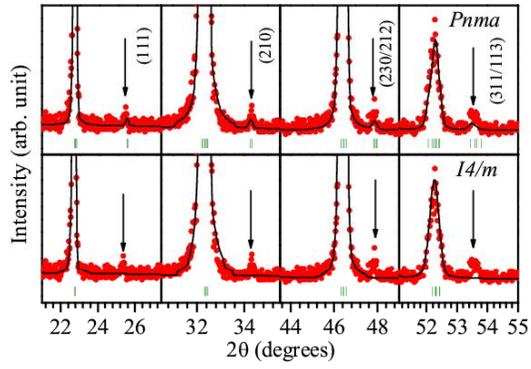

Figure 3

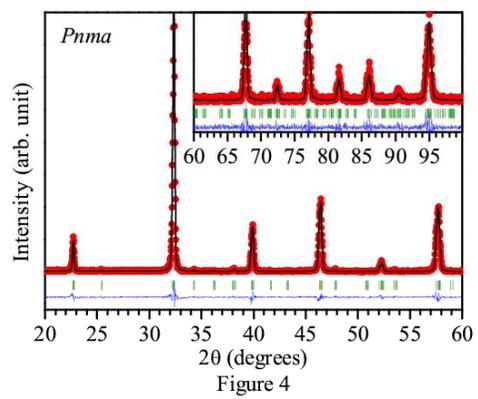

Figure 4

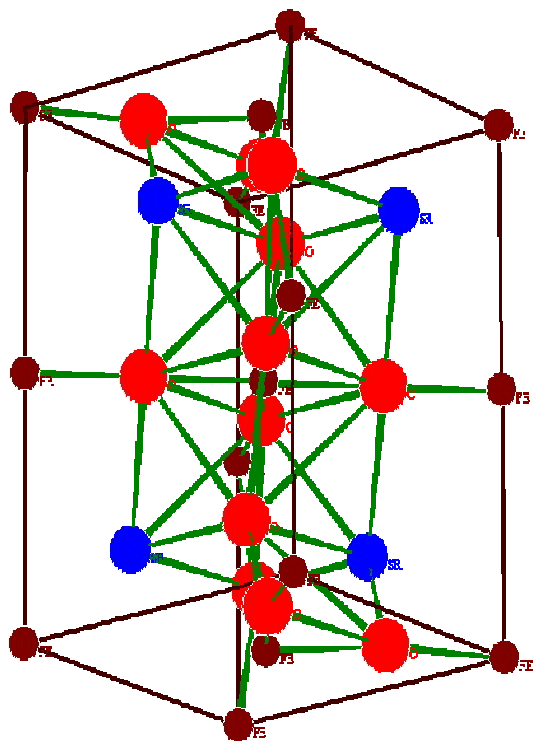

Figure 5

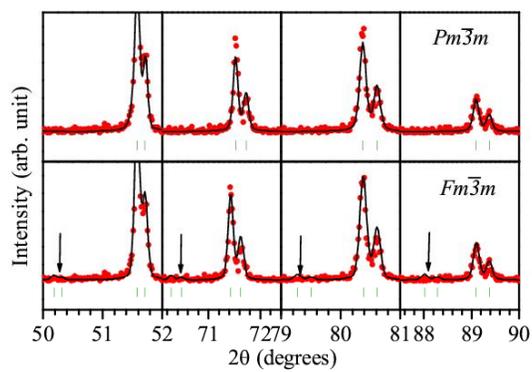

Figure 6

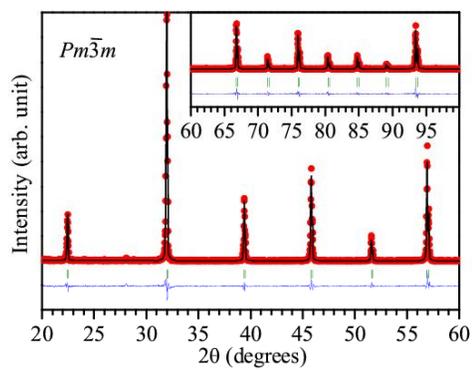

Figure 7

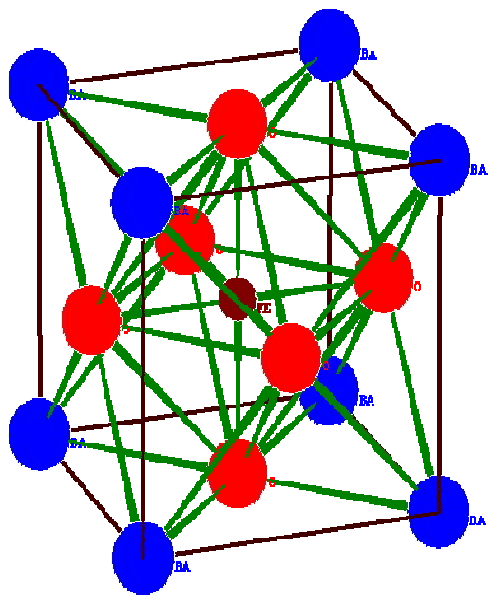

Figure 8

TABLE I

Refined structural parameters of SrLa(FeTi)O$_6$ as determined from Rietveld refinement of the orthorhombic structure in the space group *Pnma* (No.62) using x-ray powder diffraction data.

| Ions | Positional coordinates | | | Thermal parameters |
| --- | --- | --- | --- | --- |
| | x | y | Z | B(Å$^2$) |
| La/Sr | 0.0096(4) | 0.2500 | 0.500(2) | 0.74(4) |
| Fe/Ti | 0.0000 | 0.0000 | 0.0000 | 0.48(7) |
| O1 | 0.505(4) | 0.2500 | 0.575(7) | 1.08(3) |
| O2 | 0.276(3) | 0.007(4) | 0.243(7) | 0.67(2) |
| | a=5.5539(5) (Å) | b=7.8112(7) (Å) | c=5.5292(5) (Å) | V= 239.87(4) (Å$^3$) |
| | R$_{WP}$=23.16 | R$_p$= 13.96 | R$_{exp}$=19.96 | $\chi^2$= 1.36 |

TABLE II

Refined structural parameters of BaLa(FeTi)O$_6$ as determined from Rietveld refinement of the cubic structure with space group $Pm\bar{3}m$ (No.221) using x-ray powder diffraction data.

| Ions | Positional coordinates | | | Thermal parameters |
| --- | --- | --- | --- | --- |
| | x | y | Z | B(Å$^2$) |
| La/Ba | 0.0000 | 0.0000 | 0.0000 | 0.46(1) |
| Fe/Ti | 0.5000 | 0.5000 | 0.5000 | 0.28(7) |
| O | 0.5000 | 0.5000 | 0.0000 | 1.2(3) |
| | | a=3.9589(1) (Å) | | V= 62.05(8) (Å$^3$) |
| | R$_{WP}$= 25.32 | R$_p$= 15.6 | R$_{exp}$= 21.47 | $\chi^2$= 1.84 |

Table III. Selected bond lengths (Å), bond angles (°) and bond valence sum (BVS), for BaLa(FeTi)O$_6$ and SrLa(FeTi)O$_6$, obtained from Rietveld analysis of the XRD data using Cubic (*Pm$\bar{3}$m* space group) and Orthorhombic (*Pnma* space group) respectively.

|  | BaLa(FeTi)O$_6$ |
|---|---|
| **Bond distances (Å)** | |
| Fe/Ti—O(×1) | 1.9795(8) |
| Ba/La—O(×2) | 2.7994(8) |
| **Bond-Valence Sums** | |
| Ba | 2.98 |
| La | 2.20 |
| Fe | 3.30 |
| Ti | 3.85 |
| O | 2.12 |
|  | **SrLa(FeTi)O$_6$** |
| **Bond distances (Å)** | |
| Sr/La—O$_1$(×1) | 2.8330(3) |
| Sr/La—O$_1$(×1) | 2.7825(1) |
| Sr/La—O$_1$(×1) | 2.3500(4) |
| Sr/La—O$_1$(×1) | 3.1310(2) |
| Sr/La—O$_2$(×2) | 2.7949(5) |
| Sr/La—O$_2$(×2) | 2.6630(1) |
| Sr/La—O$_2$(×2) | 2.9266(3) |
| Sr/La—O$_2$(×2) | 2.6972(3) |
| Fe/Ti—O$_1$(×2) | 1.9966(1) |
| Fe/Ti—O$_2$(×2) | 2.0391(4) |
| Fe/Ti—O$_2$(×2) | 1.8894(2) |
| **Bond angles (°)** | |
| O$_1$—Fe/Ti—O$_2$ | 95.5(1) |
| O$_1$—Fe/Ti—O$_2$ | 84.3(1) |
| O$_1$—Fe/Ti—O$_2$ | 97.9(1) |
| O$_1$—Fe/Ti—O$_2$ | 82.1(1) |
| Fe/Ti—O$_1$—Fe/Ti | 156.0(1) |
| Fe/Ti—O$_2$—Fe/Ti | 171.80(2) |
| **Bond-Valence Sums** | |
| Sr | 2.27 |
| La | 2.62 |
| Fe | 3.39 |
| Ti | 3.95 |
| O1 | 2.09 |
| O2 | 2.07 |